\documentclass[3p,11pt]{elsarticle}
\usepackage{amscd}
\usepackage{amssymb}
\usepackage{amsmath}
\usepackage{stmaryrd}
\usepackage{mathrsfs}
\usepackage{times}
\usepackage{latexsym}
\usepackage{hhline}

\usepackage[latin1]{inputenc} 
\usepackage{calc}         
\usepackage{graphicx}     
\usepackage{ifthen}       
\usepackage{subfigure}    
\usepackage{pst-all}      
\usepackage{pst-poly}     
\usepackage{multido}      

\newtheorem{theorem}{Theorem}[section]
\newtheorem{lemma}[theorem]{Lemma}
\newtheorem{proposition}[theorem]{Proposition}

\newenvironment{proof}[1][Proof]{\begin{trivlist}
\item[\hskip \labelsep {\bfseries #1}]}{\end{trivlist}}

\begin{document}

\begin{frontmatter}

\title{Algebraic properties of structured context-free languages:\\ old approaches and novel developments}

\author[Milano]{Stefano Crespi Reghizzi}
\author[Milano]{Dino Mandrioli}
\address[Milano]{DEI - Politecnico di Milano, 
Piazza Leonardo da Vinci 32, \\
I-20133 Milano, Italy\\
e-mail: \{stefano.crespireghizzi, dino.mandrioli\}@polimi.it}

\begin{abstract}
The historical research line on the algebraic properties of structured CF languages initiated by McNaughton's Parenthesis Languages has recently attracted much renewed interest with the Balanced Languages, the Visibly Pushdown Automata languages (VPDA), the Synchronized Languages, and the Height-deterministic ones. Such families preserve to a varying degree the basic algebraic properties of Regular languages: boolean closure, closure under reversal, under concatenation, and Kleene star. We prove that the VPDA family is strictly contained within the Floyd Grammars (FG) family historically known as operator precedence. Languages over the same precedence matrix are known to be closed under boolean operations, and are recognized by a machine whose pop or push operations on the stack are purely determined by terminal letters. We characterize VPDA's  as the subclass of FG having a peculiarly structured set of precedence relations, and balanced grammars as a further restricted case.  The non-counting invariance property of FG has a direct implication for VPDA too.
\end{abstract}

\end{frontmatter}

\section{Introduction}
From the very beginning of formal language science, research has struggled with the wish and need to extend as far as possible the nice and powerful properties of regular languages (specifically closure properties). A major initial step has been made by McNaughton with parenthesis grammars \cite{McNaughton67}, characterized by enclosing any righthand side within a pair of parentheses; the alphabet is the disjoint union of internal letters and the pair. By considering instead of strings the stencil or skeletal trees encoded by parenthesized strings, some typical properties of regular languages that do not hold for CF languages are still valid: uniqueness of the minimal grammar, and boolean closure within the class of languages having the same rule stencils. Further mathematical developments of such ideas have been pursued in the setting of tree automata \cite{Tha67}.
\\
Several decades later, novel motivation arose for the investigation of parenthesis-like languages from the interest for mark-up languages such as XML. The \emph{balanced grammars} and languages \cite{Berstel:2001:BGT} generalize the parenthesis grammars in two ways: several pairs of parentheses are allowed, and the right-hand side of the grammar rules permit a regular expression over nonterminal and internal symbols to occur between matching parentheses. The property of uniqueness of the minimal grammar is preserved, and the family has the property of closure w.r.t. concatenation and Kleene star, that was missing in parenthesis languages. Clearly balanced as well as parenthesis languages are closed under reversal.
\\
Model checking and static program analysis provide an entirely different long-standing motivation for such families of languages --- those that extend the typical regular properties to infinite-state pushdown systems. To the best of our knowledge the seminal paper of this ``new era'' is \cite{AluMad04} which defines \emph{visibly pushdown automata} and \emph{languages} (VPDA), a subclass of realtime pushdown automata  and deterministic context-free languages. The input alphabet is partitioned into three sets named calls, returns and internals, and the decision of the type of move to perform (push, pop, or a stack neutral move) is determined by the membership of the current input letter; in other words the type of a move is solely input-driven. VPDA languages extend balanced grammars in two ways that are important for modelling symbolic program execution: they allow parentheses to remain unbalanced to represent an execution state where some procedures have not returned, and a call symbol can be matched by two or more return symbols to represent procedures with multiple exits. For each partitioned alphabet the corresponding language family is closed under the regular operations, including complement. VPDA's can be determinized and reversal produces a VPDA with calls and returns interchanged. We observe that the intended applications to static program analysis need closure under reversal in order to compute the pre- and post-reachability sets.
\\
Impulsed by this new approach, a variety of extensions and specializations of the original class have been proposed and investigated. Among them, we mention the following.
 The \emph{synchronized pushdown automata} \cite{conf/dlt/Caucal06}, instead of the fixed 3-partition of VPDA's, use
 a finite transducer that determines the type of move the PDA must perform.
\\
 The \emph{height-deterministic automata} \cite{conf/mfcs/NowotkaS07} further extended the previous idea by considering the class of PDA's characterized by the same integer-valued function returning the height of the stack for each input string; within this approach the deterministic and the real-time cases are singled out for having richer closure properties.
 Last, the \emph{synchronized grammars} \cite{caucal:DSP:2008:1743} are a more comprehensive model
 that uses an input-driven pushdown transducer to decide the type of a move. Not surprisingly, such more general models lose certain nice properties of VPL, in particular the closure under reversal, concatenation, and Kleene star.
\\
Short after McNaughton's results, we investigated similar closure properties of \emph{Floyd's} operator precedence \emph{Grammars} \cite{Floyd1963} \footnote{We propose to name them \emph{Floyd grammars} to honor the memory of Robert Floyd and also to avoid confusion with other similarly named but quite different types of precedence grammars.} (FG), an elegant precursor of LR($k$) grammars, also exploited by one of us in his work on grammar inference \cite{crespi-72}. For any given precedence matrix a syntax tree stencil is defined a priori for any word that is generated by any FG having the same precedence matrix. The family of such Floyd grammars and the related languages are a boolean algebra \cite{Crespi-ReghizziMM1978}. We also extended the notion of non-counting regular language of McNaughton and Papert \cite{McNaughtPap71} to the parenthesis languages \cite{CreGuiMan78} and to FG \cite{CreGuiMan81}.
\\
In this paper we resume the study of FG in the perspective of the cited grammatical models. We show that VPDA is a special case of FG characterized by a very restricted structure of the precedence relations, thus providing a new characterization of VPDA in terms of operator grammars. Further restrictions are shown for the case of balanced languages. Then we compare FG with the height-deterministic family showing strict inclusion, and that reversal closure is lost by that generalization.
\\
The paper is structured as follows: Section \ref{SectBasicDef} provides the essential definitions of the main classes of languages (defined through automata and/or grammars) that will be considered in this paper (others will be referred only on the basis of previous literature); Section 3 investigates the mutual inclusion relations among them. Section 4 compares the same classes of languages w.r.t their closure properties. The conclusion mentions that the non-counting invariance property of FG has a direct implication for VPDA too and shows that the whole picture of such language families deserves further analysis to answer a few remaining open issues.
\section{Basic definitions}\label{SectBasicDef}
We list the essential definitions of parenthesis and balanced grammars, VPDA, height-deterministic automata, and Floyd grammars. For brevity, other classes are not defined here because they can be somewhat put in relation with the above "basic" ones. They are nevertheless taken into consideration in Section 4. The same name is given to a class of devices (grammars or automata) and to the class of languages that can be defined by means of them.
\\
The empty string is $\varepsilon$, the terminal alphabet is $\Sigma$. For a string $x$ and a letter $a$, $|x|_a$ denotes the number of occurrences of letter $a$, and extend the notation to $|x|_{\Delta}$, for a set $\Delta\subseteq\Sigma$. Let $first(x)$ and $last(x)$ denote the first and last letter of $x\neq \varepsilon$. The projection of a string $x \in \Sigma^{\ast}$ on $\Delta$ is denoted $\pi_\Delta(x)$.
\\
The operators union, concatenation, and Kleene star are called \emph{regular}. A \emph{regular expression} is a formula written using the regular operators, parentheses and letters from a specified alphabet.
\\
A \emph{Context-Free} CF grammar is  a 4-tuple $G=(V_N, \Sigma, P, S)$, where $V_N$ is the nonterminal alphabet, $P$ the rule set, and $S$ the axiom. An \emph{empty rule} has $\varepsilon$ as the right part. A \emph{renaming rule} has one nonterminal as right part. A grammar is \emph{invertible} if no two rules have identical right parts.
\\
A rule has the \emph{operator form} if its right part has no adjacent nonterminals, and an \emph{operator grammar} (OG) contains just such rules. Any CF grammar admits an equivalent OG, which can be also assumed to be invertible \cite{Harrison78}.
\\
For a CF grammar $G$ over $\Sigma$,  the associated \emph{parenthesis grammar} \cite{McNaughton67} $\widetilde{G}$  has the rules  obtained by enclosing each right part of a rule of $G$ within the parentheses `$[$' and `$]$' that are assumed not to be in $\Sigma$.
\\
A \emph{balanced grammar} \cite{Berstel:2001:BGT} is a CF grammar  has a terminal alphabet partitioned into $\Sigma =\Sigma_{par}\cup \Sigma_i$, where $\Sigma_{par}=\{a,\overline{a},b,\overline{b}, \ldots\}$ is a set of \emph{matching parentheses} and the elements of $\Sigma_i$ are named \emph{internal}. Let $V_N$ be the nonterminal alphabet. Every rule of a balanced grammar has the form $X\to a \alpha \overline{a}$ or $X\to \alpha $, where $\alpha$ is a regular expression over $V_N\cup \Sigma_i$. The corresponding family is denoted BALAN.
\\
A \emph{pushdown automaton} PDA $\mathcal{A}$ over an alphabet $\Sigma$ is a tuple $A= (Q, \Sigma, \Gamma, \delta, q_0, F)$, where the initial state $q_0\in Q$ and $F \subseteq Q$ are the final states. $\Gamma$ is the stack alphabet containing $\bot$, the stack bottom symbol. The transition relation is
\\
$ \delta\subseteq Q \times \Gamma \times (\Sigma \cup {\varepsilon})\times Q \times (\Gamma\setminus\{\bot\})^\ast $
\\
The notation \cite{conf/mfcs/NowotkaS07} $pX \stackrel a \to q \alpha$ is equivalent to $(p, X, a, q, \alpha) \in \delta$.
\\
A PDA is called \emph{realtime} (RPDA) if $ pX \stackrel a \to q \alpha  \text{ implies } a \neq \varepsilon $.
\\
A PDA is called \emph{deterministic} (DPDA) if for every  $p\in Q, X\in \Gamma $ and $a \in \Sigma \cup \{\varepsilon\}$ we have $  |\{q\alpha \mid pX \stackrel a \to q \alpha\}| \leq  1 $ and $
  \text{if }pX \stackrel \varepsilon \to q \alpha \text{ and }pX \stackrel a \to q' \alpha' \text{ then } a= \varepsilon
$
\\
A \emph{realtime deterministic} PDA is named a RDPDA.
\\
The set $Q\Gamma^\ast$ is the set of \emph{configurations} of a PDA, with \emph{initial} configuration $q_0\bot$.
\\
The \emph{labelled transition system} generated by $\mathcal{A}$ is the edge-labeled directed graph
$$
\left( Q\Gamma^\ast\bot \;, \; \bigcup_{a \in \Sigma \; \cup \; \{\varepsilon\} } \stackrel a\longrightarrow \right)
$$
Given a string $w\in \Sigma^\ast$, we write $p \alpha \stackrel w \Longrightarrow q \beta$ if there exists a finite $w'$-labelled path, $w'\in (\Sigma  \cup \{\varepsilon\})^\ast$, from $p \alpha$ to $q \beta$, and $w$ is the projection of $w'$ onto $\Sigma$. Notice that according to \cite{conf/mfcs/NowotkaS07} the $w'$-labelled path includes transitions of the type $\stackrel \varepsilon\longrightarrow$.
\\
An $\mathcal{A}$ is \emph{complete} if $ \forall w \in \Sigma^\ast, \; q_0\bot \stackrel w \Longrightarrow q \alpha$.
\\
The language \emph{recognized} by $\mathcal{A}$ is $ L(\mathcal{A})= \{w \in \Sigma^\ast \mid q_0\bot \stackrel w \Longrightarrow p \alpha,\; p\in F\} $
\\
A PDA $\mathcal{A}$ is \emph{normalized} \cite{conf/mfcs/NowotkaS07} if
\begin{enumerate}
    \item $\mathcal{A}$ is complete;
    \item for all $p\in Q$, all rules in $\delta$ of the form $pX \stackrel a \to q \alpha$
    either satisfy $a\in \Sigma$, or all of them satisfy $a=\varepsilon$, but not both;
    \item every rule in $\delta$ is of the form
    \begin{itemize}
        \item $pX \stackrel a \to q $
        \item $pX \stackrel a \to q X $
        \item $pX \stackrel a \to q Y X $ where $a \in \Sigma \cup \{\varepsilon\} $
    \end{itemize}
\end{enumerate}
For a normalized PDA moves are named \emph{push} if $|\alpha| = 2$, \emph{pop} if $|\alpha| = 0$, and internal if $|\alpha|=1$. The normalization preserves the characteristics of DPDA, RPDA and RDPDA devices.
 \subsubsection*{Height-determinism}
 Let $w\in (\Sigma \cup \{\varepsilon\})^\ast$. The set $N(\mathcal{A},w)$ of \emph{stack heights} reached by $\mathcal{A}$ after reading $w$ is $\{|\alpha| \mid q_0\bot \stackrel w \Longrightarrow q\alpha\bot\}$. A \emph{height-deterministic} PDA (HPDA) is a PDA that is normalized and such that $|N(A,w)|\leq 1$ for every $w \in (\Sigma \cup \{\varepsilon\})^\ast$.
\\The families of height-deterministic PDA's, DPDA's, and RDPDA's  (and languages) are resp. denoted by HPDA, HDPDA, and HRDPDA.
\\
A normalized DPDA is an HDPDA and the language families HPDA and CF coincide \cite{conf/mfcs/NowotkaS07}.
\\
Two HPDA's $\mathcal{A}_1$ and $\mathcal{A}_2$ over the same alphabet $\Sigma$ are in the equivalence relation \emph{H-synchronized}, denoted by $\mathcal{A}_1\sim_H \mathcal{A}_2$, if $N(\mathcal{A}_1,w)=N(\mathcal{A}_2,w)$ for every $w \in (\Sigma \cup \{\varepsilon\})^\ast$.
\\
Let $[\mathcal{A}]_{\sim_H}$ denote the equivalence class containing the HPDA $\mathcal{A}$ and $\mathcal{A}_{HPDA}$ denote the class of languages recognized by any HPDA H-synchronized with $\mathcal{A}$.
\subsubsection*{Visibly pushdown automata}
A \emph{visibly pushdown} (VP) \cite{AluMad04} alphabet is a 3-tuple $\widehat{\Sigma}= \langle\Sigma_{c}, \Sigma_{r},\Sigma_{i}\rangle$, with $\Sigma$ the disjoint union of the three sets.  Elements of the three sets are resp. termed \emph{calls}, \emph{returns} and \emph{internal} letters.
\label{DetvisibPushdownAut} A  \emph{VP automaton} VPDA is a  PDA $\mathcal{A}=(\Sigma,Q, q_0, \Gamma, \delta, F)$, where $\widehat{\Sigma}$ is a VP alphabet. The transition relation is
$$
\delta\subseteq \left(Q \times \Sigma_{c} \times Q \times (\Gamma \setminus \{\bot\}\right) \quad \cup\quad \left(Q \times \Sigma_{r} \times \Gamma \times Q \right) \quad\cup \quad\left( Q \times \Sigma_{i} \times Q \right)
$$
that can be readily seen to specialize the previous definition for a general PDA.
\subsubsection*{Floyd or operator precedence grammars}
The  definitions for operator precedence grammars, here renamed \emph{Floyd Grammars} (FG), are from \cite{Crespi-ReghizziMM1978}. (See also \cite{GruneJacobs:08} for a recent presentation.) \label{LeftRightSets}
\\
For a nonterminal $A$ of an OG $G$, the \emph{left and right terminal sets} are
$$
  \mathcal{L}_G(A) = \{a\in\Sigma \mid A \stackrel \ast \Rightarrow B a\alpha\} \qquad
   \mathcal{R}_G(A) = \{a\in\Sigma \mid A \stackrel \ast \Rightarrow \alpha a B\}
$$
where $B\in V_N\cup\{\varepsilon\}$ and $\stackrel \ast \Rightarrow$ denotes, as usual, a derivation. The two definitions are extended to a set $W$ of nonterminals and to a string $\beta \in V^+$ via
$$
\mathcal{L}_G(W)=\bigcup_{A\in W}\mathcal{L}_G(A)
 \text{ and }
    \mathcal{L}_G(\beta)=\mathcal{L}_{G'}(D) \\
$$
 where $D$ is a new nonterminal and $G'$ is the same as $G$ except for the addition of the
 rule $D\to \beta$.  Finally  $\mathcal{L}_G(\epsilon)=\emptyset $. The definitions for $\mathcal{R}$ are similar.
\\
 \label{PrecRelat}
For an OG $G$, let $\alpha, \beta \in (V_N \cup \Sigma)^{\ast}$ and  $a,b\in \Sigma$, three binary operator precedence (OP) relations are defined:
\begin{center}
\begin{tabular}{cc}
  equal precedence:  & $a\doteq b $ iff $ \exists A\to\alpha aBb\beta$, $B\in V_N\cup\{\varepsilon\}$; \\
  yields precedence:  & $a\gtrdot  b $ iff $ \exists A\to\alpha Db\beta$, $D\in V_N$ and $a\in \mathcal{R}_G(D)$ \\
  takes precedence: & $a\lessdot  b $ iff $ \exists A\to\alpha aD\beta$, $D\in V_N$ and $b\in \mathcal{L}_G(D)$; \\
\end{tabular}
\end{center}
\label{OPMatrixOPG} For an OG $G$, the \emph{operator precedence matrix} (OPM) $M=OPM(G)$ is a $|\Sigma| \times |\Sigma|$ array that to each ordered pair $(a,b)$ associates the set $M_{ab}$ of OP relations holding between $a$ and $b$.
Given two  OPM's  $M_1$ and $M_2$, we define
\\
$
  M_1\subseteq M_2 \iff M_{1,ab}\subseteq M_{2,ab}, \quad
  M=M_1\cup M_2  \iff  M_{ab}= M_{1,ab}\cup M_{2,ab};  \forall a,b.
$
\\
$G$ is a \emph{Floyd grammar} FG if, and only if, $OPM(G)$ is a \emph{conflict-free} matrix, i.e., $\forall a,b$, $|OPM(G)_{ab}|\leq 1$. Two matrices are \emph{compatible} if their union is conflict-free.
\\
A FG is in \emph{Fischer normal form \cite{Fischer69}} if it is invertible, the axiom $S$ does not occur in the right part of any rule, and there are no renaming rules, except those with left part $S$ (if any).
\par
For the reader convenience the acronyms are collected in the table:
\begin{center}
\begin{tabular}{l|l}
  \hline
  BALAN & balanced grammar \\
  CF & context-free \\
  DPDA & deterministic pushdown automaton \\
  FG    & Floyd grammar \\
  HDPDA & height-deterministic deterministic pushdown automaton \\
  HPDA & height-deterministic  pushdown automaton \\
  HRDPDA & height-deterministic realtime deterministic pushdown automaton \\
  OG & operator grammar \\
  OPM & operator precedence matrix \\
  REG & regular language \\
  RDPDA & realtime deterministic pushdown automaton \\
    RPDA & realtime pushdown automaton \\
  PDA & pushdown automaton \\
  VPDA & visibly pushdown automaton\\
  \hline
\end{tabular}
\end{center}
\section{Containment relations}
First we recall some of the relevant known \cite{AluMad04,conf/mfcs/NowotkaS07,conf/csr/LimayeMM08,caucal:DSP:2008:1743} containment relations between some recent language families, then we position FG within the picture. The main  strict inclusions are:
\begin{center}
$ REG \subset BALAN \subset VPDA \subset  HRDPDA=RDPDA \subset  HDPDA = DPDA $
\end{center}
Notice that the above inclusions preserve the structural properties of the languages: for instance if the partition of a VP alphabet places a letter in $\Sigma_c$ and therefore associates a push move to it, the corresponding HDPDA automaton too performs a push move on that letter.
\\
The first \cite{conf/dlt/Caucal06} and second \cite{conf/csr/CaucalH08} family of Caucal, as well as the one of Fisman and Pnueli \cite{FisPnu01} fall in between VPDA and DPDA. but lack of space prevents a detailed presentation.
\\
\label{SectVPAareOPG} Next we focus on FG  languages. It is well-known that $FG \subset DPDA$. On the other hand, FG includes non-realtime deterministic languages such as $L_1=\{a^m b^n c^n d^m \mid m,n\ge 1\} \cup \{a^m b^+ e d^m \mid m\ge 1\}$. Observing that $L_2=\{a^n c a^n \mid n\geq 0\}$ is in HRDPDA but not in FG, since, by an elementary application of the pumping lemma, this would imply a precedence conflict, we have:
 \begin{proposition}
 The families of FG and HRDPDA languages are incomparable.
 \end{proposition}
 Our main result is that the VPDA languages are a well-characterized special case of
 FG languages. First we give a construction from a VPDA to a FG having a
 certain type of precedence matrix, second we construct a VPDA for any FG with such matrices.
At last we include also BALAN in the matrix-based characterization.
\\
 We need to analyze the structure of  VPDA strings.
 A string in $\{c,r\}^{\ast}$ is \emph{well parenthesized} if it reduces to $\varepsilon$ via the cancellation rule $cr\to \varepsilon$.
\\
Let $\rho$ be the alphabetical mapping from $\Sigma_c \cup \Sigma_r\cup \Sigma_i $ to $\{c,r\}$ defined by $\rho(c_j)=c, \forall c_j\in \Sigma_c$, $\rho(r_j)=r, \forall r_j\in \Sigma_r$, and $\rho(s_j)=\varepsilon, \forall s_j\in \Sigma_i$. A non-empty string $x \in \Sigma^{\ast}$ is \emph{well balanced} if  $\rho(y)$ is well parenthesized; it is \emph{well closed} if in addition $first(x)\in \Sigma_c$ and $last(x)\in \Sigma_r$.
\\
Let $\mathcal{A}=(Q, \Sigma,Q, q_0, \Gamma,  \delta, Q_F)$ be a VPDA, with $\Sigma=\Sigma_{c}\cup \Sigma_{r} \cup \Sigma_{i}$.
\begin{lemma}\label{LemmaStruttVPL}
 Any
string $x \in L(\mathcal{A})$ can be factorized as\\
$ x = y c_0 z$   or $x = y$,   with $c_0\in \Sigma_c$, such that
\begin{enumerate}
\item $ y=u_1 w_1 u_2 w_2 \ldots u_k w_k, k\geq 0, $ where $u_j \in (\Sigma_i \cup \Sigma_r)^\ast$, and $w_j\in \Sigma^\ast$ is a, possibly missing, well-closed string;
    \item
    $
 z = v_1 c_1 v_2 c_2 \ldots c_{r-1} v_r,\qquad r\geq 0,
    $
 where $c_j \in \Sigma_c$ and $v_j \in \Sigma^\ast$ is a, possibly null, well-balanced string.
\end{enumerate}
\begin{proof}
 Let the transitions from state $q$ to $q'$  be labelled as follows: $(r,\bot)$ denotes a move of type $(q,r,\bot,q')\in\delta_r$;
 $(r,Z)$  denotes a move of type $(q,r,Z,q')\in\delta_r$ with $Z\neq \bot$;
 $\frac{c}{Z}$  denotes a move of type $(q,c,q',Z)\in\delta_c$;
 $s$  denotes  a move of type $(q,s,q')\in\delta_s$.
 \\
  We examine the possible sequences of moves of a suitable VPDA $\mathcal{A}$ that for convenience is non-deterministic (determinization is always possible \cite{AluMad04}).
  We only discuss the case $x = y c_0 z$, since the case $x = y$ is simpler.
\\
The computation starts with a series of moves in $\{(r,\bot)\mid s\}^\ast$, which scan the prefix $u_1$
 and leave the stack empty.
\\ Then the machine may do a series of moves to scan string $w_1$.
    The first move is of type $\frac{c}{Z_i}$. The move is
 possibly followed  by a nested computation scanning a well-balanced string, and at last by a move of type $(r,Z_i)$. The effect is  to scan a well balanced string $w_1$. Clearly the nested computation may also include internal moves.
\\ After scanning $w_1$ the stack is empty, and the
 computation may scan  $u_2$, and so on, until $w_k$ is scanned.
\\ Alternatively and non-deterministically, when the stack is empty, the machine may perform a move
    $\frac{c_0}{Z_U}$,  thus  entering the phase that scans string $z$. We denote as $Z_U$ a symbol written on the stack, which will never be touched by a subsequent pop move. In other words, $c_0$ is nondeterministically assumed to be an unmatched call.
\\ Then the $z$ phase non-deterministically scans a well balanced string $v_{1}$. Then, again nondeterministically, it may
perform a move $\frac{c_1}{Z_U}$. Then it may scan another well balanced string $v_{2}$, and so on, ending with a stack in $\bot {Z_U}^+$.
\\
  At any time, when the machine enters a final state, it may halt and recognize the scanned input.
\end{proof}
\end{lemma}
Clearly  string $y$ is the longest prefix such that the accepting computation ends with empty stack. For simplicity, without loss of generality, we assume that no transition enters the initial state $q_0$. For convenience we shall denote by a subscripted letter $q$ the states traversed while scanning $y$, and by a subscripted letter $p$ the states traversed in the computation of $c_0z$. The state set is thus partitioned into $Q= \{q_0\}\cup Q_q \cup Q_p$.
\par
\begin{figure}[h!]
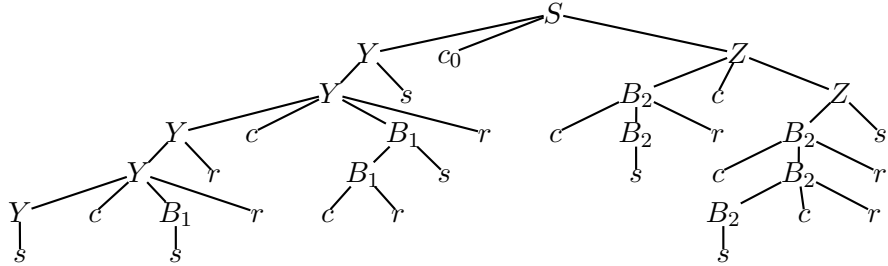

\begin{center}
 \scalebox{1.0}{%
\psset{arrows=-,linestyle=solid,levelsep=15pt}

\pstree{\TR{$S$}}
 {
  \pstree{\TR{$Y$}}
    {

   \pstree{\TR{$Y$}}
    {

   \pstree{\TR{$Y$}}
    {

   \pstree{\TR{$Y$}}
    {

    \pstree{\TR{$Y$}}
    {
    \TR{$s$}
    }

    \TR{$c$}

    \pstree{\TR{$B_1$}}
    {
    \TR{$s$}
    }

    \TR{$r$}

    }
    \TR{$r$}

    }
    \TR{$c$}

    \pstree{\TR{$B_1$}}
    {
    \pstree{\TR{$B_1$}}
    {
    \TR{$c$}  \TR{$r$}
    }

    \TR{$s$}
    }
    \TR{$r$}
    }
    \TR{$s$}
    }

 \TR{$c_0$}

      \pstree{\TR{$Z$}}
    {

      \pstree{\TR{$B_2$}}
    {
    \TR{$c$}

    \pstree{\TR{$B_2$}}
    {
    \TR{$s$}
    }

    \TR{$r$}

    }

    \TR{$c$}
    \pstree{\TR{$Z$}}
      {
    \pstree{\TR{$B_2$}}
    {
    \TR{$c$}
   \pstree{\TR{$B_2$}}
    {
    \pstree{\TR{$B_2$}}{\TR{$s$}}

    \TR{$c$}\TR{$r$}

    }
    \TR{$r$}

    }
    \TR{$s$}

      }

    }

 }
 }
\end{center}
\caption{\label{figureTree}Schema of a syntax tree generated by the precedence grammar constructed in Theor. \ref{TheorVPD2OPG}.}
\end{figure}

Since VPL's are CF languages, previous papers (e.g. \cite{LaTorreNP06}) have also used grammars to define them,
 but such grammars are not OG or have precedence conflicts; instead, we present a construction producing a grammar with the required properties.
\begin{theorem}\label{TheorVPD2OPG}
For any  visibly pushdown automaton $\mathcal{A}$ a Floyd grammar $G$ such that $L(G)=L(\mathcal{A})$ can be effectively constructed.
\begin{proof}
 First we construct the grammar, then we prove that it is an FG, and lastly that it is equivalent to $\mathcal{A}$.
\end{proof}
\end{theorem}
\emph{Grammar construction}.\label{OPGgrammarConstr}
\\
 The rules are keyed to the factorization of Lemma \ref{LemmaStruttVPL} and are listed in Tables \ref{productionsS}, \ref{productionsY},  \ref{productionsB}, and \ref{productionsZ}. The scheme of a sample syntax tree produced by the grammar, for a string factorized as in Lemma \ref{LemmaStruttVPL}, is shown in Fig. \ref{figureTree}.

\begin{figure}[h]
\begin{center}
$
\begin{array}{c|c|c|c|}
    & \Sigma_c & \Sigma_r& \Sigma_i  \\\hline
 \Sigma_c  & \lessdot  & \doteq &  \lessdot \\\hline
 \Sigma_r  & \gtrdot  & \gtrdot  & \gtrdot  \\\hline
  \Sigma_i   & \gtrdot  & \gtrdot  & \gtrdot  \\
  \hline
\end{array}
$
\end{center}
\caption{\label{figureTotalVPmatr}\emph{Total VP precedence matrix} $M_T$. }
\end{figure}
Nonterminals of class $Y$ generate a string such that the automaton, parsing it, starts and ends with empty stack. Nonterminals of classes $B_1,B_2$ derive a well-balanced (but not necessarily well-closed) string. Nonterminals of class $Z$ derive a string such that, starting with a non-empty stack of the form $\bot Z_U^+$, the stack never pops a $Z_U$ and at last contains a string in $\bot Z_U^+$.
\\
The nonterminal symbols of the grammar are denoted  by a pair of states $\langle q_i, q_j\rangle$ or $\langle p_i, p_j\rangle$, or by a triple $\langle q_i,Z,q_j\rangle$ or $\langle p_i,Z,p_j\rangle$, with $Z \in \Gamma$. Intuitively, a nonterminal of the generic form $\langle r_i\ldots r_j\rangle$ generates a terminal string $u$ if, and only if, there is a computation of the machine from the left state $r_i$ to the right state  $r_j$ which reads the same string and never modifies the initial stack. Furthermore, nonterminals $\langle q_i, q_j\rangle$ leave the stack unchanged; nonterminals $\langle p_i, p_j\rangle$ at most increase the number of $Z_U$'s; and nonterminals $\langle q_i,Z,q_j\rangle$ or $\langle p_i,Z,p_j\rangle$ denote that the computation starts and ends with $Z$ on the top and generates a well-balanced terminal string $w$.
\\
To construct the rules we examine the transitions of the  VPDA. In what follows, calls, returns and internal letters are respectively denoted $c, r$ and $s$; $Z,W$ are stack symbols different from $\bot$.
\begin{table}[h!]
\caption{\label{productionsS}Productions of the axiom. }
\renewcommand{\arraystretch}{1.3}
\begin{tabular}{l|p{5cm}|p{6cm}}
  case & transitions  & rules \\
  \hline
 $S\to Yc_0 Z$ & $\delta(q_i, c_0)\ni(p_j, Z_U)$ &   $S \to \langle q_0,q_i\rangle c_0 \langle p_j, p_f\rangle$, $\forall p_f \in F$ \\

   $S\to Y$   &   &  $S \to \langle q_0,q_f\rangle$, $\forall q_f \in F$  \\

    $S\to Y c_0 $   & $\delta(q_i, c_0)\ni(p_f, Z_U)$, $p_f \in F$&   $S \to \langle q_0,q_i\rangle c_0$ \\

    $S\to  c_0 Z$    & $\delta(q_0, c_0)\ni(p_j, Z_U)$ &  $S \to  c_0 \langle p_j, p_f\rangle$, $\forall p_f \in F$ \\

   $S\to  c_0$   & $\delta(q_0, c_0)\ni(p_f, Z_U)$, $p_f \in F$ &  $S \to  c_0$\\
  \hline
\end{tabular}
\end{table}
\begin{table}[h!]
\caption{\label{productionsY}Productions of nonterminals of class $Y$ (deriving the maximal prefix ending with empty stack). }
\renewcommand{\arraystretch}{1.3}
\begin{tabular}{l|p{6cm}|l}
  case & transitions  & rules \\
  \hline
 $Y\to s$ & $\delta(q_0, s)\ni q_i$ &   $\langle  q_0,q_i\rangle\to  s$  \\

 $Y\to r$    & $\delta(q_0, r, \bot)\ni q_i$ &  $\langle q_0,q_i\rangle\to  r$ \\

 $Y\to Y s$    & $\delta(q_i, s)\ni q_j$ &  $\langle q_0,q_j\rangle\to \langle q_0,q_i\rangle s$ \\

 $Y\to Y r$    & $\delta(q_i, r, \bot)\ni q_j$ & $\langle q_0,q_j\rangle\to \langle q_0,q_i\rangle r$\\

 $Y\to c B r$   & $\delta(q_0, c)\ni (q_t, Z)$ and  $\delta(q_k, r, Z)\ni q_h $ & $\langle q_0, q_h\rangle\to c \langle q_t, Z, q_k\rangle r$\\

$Y\to c r$    & $\delta(q_0, c)\ni (q_t, Z)$ and  $\delta(q_t, r, Z)\ni q_h $ & $\langle q_0, q_h\rangle\to c r$\\

 $Y\to Y c B r$ & $\delta(q_i, c)\ni (q_j, Z)$ and $\delta(q_m, r, Z)\ni q_n$ & $\langle q_0,q_n \rangle \to \langle q_0,q_i \rangle c \langle q_j, Z, q_m\rangle r$\\

 $Y\to Y c  r$  & $\delta(q_i, c)\ni (q_j, Z)$ and $\delta(q_m, r, Z)\ni q_n$ and $q_j=q_m$ & $\langle q_0,q_n \rangle \to \langle q_0,q_i \rangle c  r $ \\

  \hline
\end{tabular}
\end{table}
\begin{table}[h!]
\caption{\label{productionsB}Productions for  nonterminals of classes $B_1$ and $B_2$, generating well-balanced string. (The case $B_2$ just differs with respect to the state set, which is $Q_p$ instead of $Q_q$.) }
\renewcommand{\arraystretch}{1.3}
\begin{tabular}{l|p{6,3cm}|p{7,0cm}}
  case & transitions  & rules \\
  \hline
 $ B\to BcBr$ & $\delta(q_i, c)\ni (q_j, Z)$ and $\delta(q_m, r, Z)\ni q_n$
                   & $\langle q,q_n \rangle  \to \langle q,q_i \rangle c \langle q_j, Z, q_m\rangle r$, ${\forall q \in Q_q}$\\

 $B\to Bcr$   &   $\delta(q_i, c)\ni (q_j, Z)$ and $\delta(q_j, r, Z)\ni q_n$
                        & $\langle q,q_n \rangle  \to \langle q,q_i \rangle c  r $, ${\forall q \in Q_q}$ \\

$B\to cBr$   &   $\delta(q_i, c)\ni (q_j, Z)$ and $\delta(q_m, r, Z)\ni q_n$
                        & $\langle q_i,q_n \rangle  \to c \langle q_j, Z,q_n \rangle r $ \\

 $B\to cr$   & $\delta(q_i, c)\ni (q_j, Z)$ and $\delta(q_j, r, Z)\ni q_n$
                      & $\langle q_i,q_n \rangle  \to  c  r $, ${\forall q \in Q_q}$ \\

$B\to BcBr$ & $\delta(q_i, c)\ni (q_j, Z)$ and $\delta(q_m, r, Z)\ni q_n$
                   & ${\langle q_i,W,q_n \rangle  \to \langle q,q_i \rangle c \langle q_j, Z, q_m\rangle r}$, ${\forall q \in Q_q, W\in\Gamma}$ \\

$B\to cBr$   &   $\delta(q_i, c)\ni (q_j, Z)$ and $\delta(q_m, r, Z)\ni q_n$ & $\langle q,W,q_n \rangle \to c \langle q_j,Z,q_m \rangle r$, ${\forall q \in Q_q,W\in\Gamma}$  \\

 $B\to Bcr$ & $\delta(q_i, c)\ni (q_j, Z)$ and $\delta(q_j, r, Z)\ni q_n$  & $\langle q,W,q_n \rangle \to \langle q,q_i \rangle c r$, ${\forall q \in Q_q,W\in\Gamma}$  \\

 $B\to Bs$   & $\delta(q_h, s)\ni q_m$ & $\langle q, W, q_m \rangle \to  \langle q, q_h \rangle s$, ${\forall q \in Q_q, W\in\Gamma}$ \\

  $B\to s$    & $\delta(q_j, s)\ni q_m$ & $\langle q_j, Z, q_m \rangle \to s$, ${\forall W\in\Gamma}$ \\\hline
\end{tabular}
\end{table}
\begin{table}[h!]
\caption{\label{productionsZ}Productions for  nonterminals of class $Z$. }
\renewcommand{\arraystretch}{1.3}
\begin{tabular}{l|p{6,2cm}|p{7cm}}
  case & transitions  & rules \\
  \hline
 $Z\to c Z$ & $\delta(p_i, c)\ni (p_j,Z_U)$ & $\langle p_i, p_f\rangle\to \ c \langle p_j, p_f\rangle$, ${\forall p_f \in F}$ \\

 $Z\to c$ & $\delta(p_i, c)\ni (p_f,Z_U)$, $p_f \in F$ &  $\langle p_i, p_f\rangle\to c$\\

 $Z\to B c Z$      & $\delta(p_j, c)\ni (p_h,Z_U)$ & $\langle p, p_f\rangle\to \langle p, p_j\rangle c \langle p_h, p_f\rangle$, ${\forall p_f\in F, p\in Q_p}$
     \\

  $Z\to B C$    & $\delta(p_j, c)\ni (p_f,Z_U)$, $p_f \in F$  & $\langle p, p_f\rangle\to \langle p, p_j\rangle c$ \\

   $Z\to BcBr$ & $\delta(p_i, c)\ni (p_j, Z)$ and $\delta(p_m, r, Z)\ni p_n$
                   & $\langle p,p_f \rangle  \to \langle p,p_i \rangle c \langle p_j, Z, p_m\rangle r$, ${\forall p \in Q_p,  p_f\in F}$\\

   $Z\to cBr$   &   $\delta(p_i, c)\ni (p_j, Z)$ and $\delta(p_m, r, Z)\ni p_n$
                        & $\langle p_i,p_f \rangle  \to c \langle p_j, Z,p_n \rangle r $, ${\forall p_f\in F}$\\

   $Z\to Bcr$   &   $\delta(p_i, c)\ni (p_j, Z)$ and $\delta(p_j, r, Z)\ni p_n$
                        & $\langle p,p_f \rangle  \to \langle p,p_i \rangle c  r $, ${\forall p \in Q_p}$ , $\forall p_f\in F$\\

  $Z\to cr$   & $\delta(p_i, c)\ni (p_j, Z)$ and $\delta(p_j, r, Z)\ni p_n$
                      & $\langle p_i,p_f \rangle  \to  c  r $, ${\forall p \in Q_p, p_f\in F}$ \\

   $Z\to B s$    &  $\delta(p_j, s)\ni p_f$, $p_f \in F$    &   $\langle p, p_f\rangle\to \langle p, p_j\rangle s$, ${\forall p\in Q_p}$  \\

   $Z\to s$    &   $\delta(p_j, s)\ni p_f$ , $p_f \in F$        &   $\langle p, p_f\rangle\to  s$ \\
\hline
\end{tabular}
\end{table}
 Notice that the grammar  constructed may be not reduced (i.e. some nonterminal may be unreachable from the axiom or it may not derive any terminal string). In that case the useless nonterminals and rules can be removed by well-known algorithms (e.g. in \cite{hopullman:automata}).
\subsubsection*{$G$ is a Floyd grammar}
By construction all the rules are in operator form. To verify that the operator precedence matrix $M$ is conflict-free, it suffices to compute  the relevant terminal sets the matrix entries using  the previous definitions. It should be enough to show one case.
\\
 For the rule $\langle q_0,q_n \rangle\to \langle q_0,q_i \rangle c \langle q_j,Z,q_m \rangle r$ the set
$ \mathcal{R}_G(\langle q_0,q_i \rangle)\subseteq \Sigma_i \cup  \Sigma_r $ produces the relations $s\gtrdot c, r\gtrdot c$. The sets $ \mathcal{L}_G(\langle q_j,Z, q_m \rangle)\subseteq \Sigma_i \cup \Sigma_c$, $\mathcal{R}_G(\langle q_j,Z, q_m \rangle)\subseteq \Sigma_i \cup \Sigma_r$  determine $c\lessdot c, c\lessdot s$ and $s\gtrdot r, r\gtrdot r$; the right part of the rule gives $c\dot=r$. Thus we obtain a conflict-free matrix $M\subseteq M_{T}$ where $M_{T}$ is the total matrix in Fig. \ref{figureTotalVPmatr}.

\begin{figure}[h]
\begin{center}
\scalebox{0.84}{%
\psset{arrows=->,labelsep=1pt,colsep=1mm, rowsep=3mm}
\begin{psmatrix}
\\
$\vdash$ & - & - & - & - & - & - & - & - & - & - & - & - & - & - & - & - & - & - & - & - & - & - & - & - & - & - & - & - & - & - & - & - & - & - & - & - & - & - &- & - & - & - & - & - & - & $\dashv$
\\
$\vdash$ & $\lessdot $ & - & - & - & - & - & - & - & - & - & - & $c $ & $\gtrdot $ & - & - & - & - & - & - & - & - & - & - & - & - & - & - & - & - & - & - & - & - & - & - & - & - & - &- & - & - & - & - & - & - & $\dashv$
\\
$\vdash$ & $\lessdot $ & - & - & - & - & - & - & - & - & - & - & $c $ & $\lessdot $ & - & - & - & - & - & - & - & - & - & - & $c_0$ & $\gtrdot $ & - & - & - & - & - & - & - & - & - & - & - & - & - &- & - & - & - & - & - & - & $\dashv$
\\
$\vdash$ & $\lessdot $ & - & - & - & - & - & - & - & - & - & - & $c $ & $\lessdot $ & - & - & - & - & - & - & - & - & - & - & $c_0$ & $\lessdot $ & - & - & - & - & - & - & $c $ & $\gtrdot $ & - & - & - & - & - &- & - & - & - & - & - & - & $\dashv$
\\
$\vdash$ & $\lessdot $ & - & - & - & - & - & - & - & - & - & - & $c $ & $\lessdot $ & - & - & - & - & - & - & - & - & $s$ & $\gtrdot $ & $c_0$ & $\lessdot $ & - & - & - & - & - & - & $c $ & $\lessdot $ & - & - & - & - & - &- & - & - & - & - & $s$ & $\gtrdot $ & $\dashv$
\\
$\vdash$ & $\lessdot $ & - & - & - & - & - & - & - & - & $r$ & $\gtrdot $ & $c $ & $\lessdot $ & - & - & - & - & - & - & $r$ & $\gtrdot $ & $s$ & $\gtrdot $ & $c_0$ & $\lessdot $ & - & - & - & - & - & - & $c $ & $\lessdot $ & $c$ & $\dot=$ & - & - & - &- & - & - & $r$ & $\gtrdot $ & $s$ & $\gtrdot $ & $\dashv$
\\
$\vdash$ & $\lessdot $ & - & - & $c$ & $\dot=$ & - & - & $r$ & $\gtrdot $ & $r$ & $\gtrdot $ & $c $ & $\lessdot $ & - & - & - & - & $s$ & $\gtrdot $ & $r$ & $\gtrdot $ & $s$ & $\gtrdot $ & $c_0$ & $\lessdot $ & $c$ & $\dot=$ & - & - & $r$ & $\gtrdot $ & $c $ & $\lessdot $ & $c$ & $\lessdot $ & - & - & $c$ & $\dot=$ & $r$ & $\gtrdot $ & $r$ & $\gtrdot $ & $s$ & $\gtrdot $ & $\dashv$
\\
$\vdash$ & $\lessdot $ & $s$ & $\gtrdot $ & $c$ & $\lessdot $ & $s$ & $\gtrdot $ & $r$ & $\gtrdot $ & $r$ & $\gtrdot $ & $c $ & $\lessdot $ & $c $ & $\dot=$ & $r$ & $\gtrdot $ & $s$ & $\gtrdot $ & $r$ & $\gtrdot $ & $s$ & $\gtrdot $ & $c_0$ & $\lessdot $ & $c$ & $\lessdot $ & $s$ & $\gtrdot $ & $r$ & $\gtrdot $ & $c $ & $\lessdot $ & $c$ & $\lessdot $ & $s$ & $\gtrdot $ & $c$ & $\dot=$ & $r$ & $\gtrdot $ & $r$ & $\gtrdot $ & $s$ & $\gtrdot $ & $\dashv$
 \\

%

\end{psmatrix}
}

$ \vdash \lessdot  \underbrace{s}_{u_1} \gtrdot  \underbrace{c \lessdot  s \gtrdot  r}_{w_1} \gtrdot \underbrace{r}_{u_2} \gtrdot  \underbrace{c \lessdot c \dot=r \gtrdot s\gtrdot r}_{w_2} \gtrdot \underbrace{s}_{u_3} \gtrdot c_0\lessdot  \underbrace{c\lessdot s\gtrdot r}_{v_1}\gtrdot  c \lessdot \underbrace{c\lessdot  s\gtrdot c\dot=r \gtrdot r \gtrdot s}_{v_2} \gtrdot \dashv $
\end{center}
\caption{\label{figureParsingRelations}Precedence relations between letters during the parsing of the string of Fig. \ref{figureTree}. The dummy string delimiters $\vdash, \dashv$ by hypothesis respectively yield and take precedence over any other letter. }
\end{figure}
Fig. \ref{figureParsingRelations} reproduces the string of Fig. \ref{figureTree} with precedence relations between letters that are consecutive or separated by a nonterminal.
\subsubsection*{Proof that $L(G)= L(\mathcal{A})$}
It is obtained by a fairly natural induction showing the double implication between computations and derivations. It is structured into several ``macro-steps'' mirroring the factorization introduced in Lemma \ref{LemmaStruttVPL}. We develop in detail only a sample of the various cases, since the others are similar.
\begin{enumerate}
    \item $(q_i, \bot)  \underset{x}{\stackrel*\mapsto} (q_j,  \bot) \iff
    \langle  q_i,q_j \rangle \stackrel{\ast}{\Rightarrow} x$, $x \in (\Sigma_r \cup \Sigma_i)^\ast$.
    \\

    \item $(q_i, \sigma)  \underset{x}{\stackrel*\mapsto} (q_j,  \sigma) \iff
    \langle  q_i,q_j \rangle \stackrel{\ast}{\Rightarrow} x$, $x \in  \Sigma^\ast$ and well-balanced.
    \\

    \item $(p_i, \sigma)  \underset{x}{\stackrel*\mapsto} (p_j, \sigma) \iff
    \langle p_i, p_j\rangle \stackrel{\ast}{\Rightarrow} x$, $x \in \Sigma^\ast$ and well-balanced.
    \\

    \item $(p_i, \bot Z_U^k)  \underset{c^n}{\stackrel*\mapsto} (p_j, \bot Z_U^{k+n})\iff
    \langle p_i, p_j\rangle \stackrel{\ast}{\Rightarrow} c^n$.
    \\
    \item $\forall \gamma \in \Gamma^\ast, Z$,
    $(p_i, \bot \gamma Z)  \underset{w}{\stackrel*\mapsto} (p_j,\bot \gamma Z)$ (without ever popping $Z$)
    $ \iff
    \langle p_i, Z, p_j\rangle \stackrel{\ast}{\Rightarrow} w$, where $w$ is a well-balanced string.
    \\
    Induction base:
    \begin{enumerate}
        \item $\delta(p_i,c)\ni (p_k, Z) \wedge \delta(p_k,r,Z)\ni p_r \iff
       \exists W:  \langle p_i, W, p_j\rangle \to cr$

        \item $\delta(p_i,s)\ni p_j \iff \exists W:  \langle p_i, W, p_j\rangle \to s$
    \end{enumerate}
     From the inductive hypotheses:
     \begin{enumerate}
        \item $(p_i, \bot \gamma W)  \underset{x}{\stackrel*\mapsto} (p_h, \bot \gamma W)\iff
         \langle p_i, p_h\rangle \stackrel{\ast}{\Rightarrow} x, x\in \Sigma^\ast$

        \item  $(p_h, \bot \gamma W)  \underset{c}{\mapsto} (p_t, \bot \gamma WZ)$

        \item  $(p_t, \bot \gamma WZ)\underset{w_1}{\stackrel*\mapsto} (p_r, \bot \gamma WZ)\iff
            \langle p_t,Z, p_r\rangle \stackrel{\ast}{\Rightarrow} w_1$

        \item  $(p_r, \bot \gamma W Z)  \underset{r}{\mapsto} (p_j, \bot \gamma W)$

    \end{enumerate}
\end{enumerate}
we derive:
\begin{equation}\label{lastImplication}
 (p_i, \bot \gamma W)  \underset{w}{\stackrel*\mapsto}  (p_j, \bot \gamma W) \iff
 \langle p_i,W, p_j\rangle \stackrel{\ast}{\Rightarrow} w, w = xc w_1 r
\end{equation}
Special cases, such as $x=\varepsilon$ and many others, can be similarly treated.
 \qed
 N.B. Each inductive proof of the various assertions may exploit other assertions in the inductive steps.
 For instance the inductive hypothesis (a) above is based on assertion 3.
\par
\label{SectStrictInclus} A natural question is whether every FG defines a VPDA language or not.
\begin{theorem}\label{TheorStrictInclusion}
The VPDA language family is strictly included in the FG family.
\begin{proof}
\label{ProposCounterexample} The  language
$$
L=\{b^n c^n \mid n\geq 1\}\cup \{f^n d^n \mid n\geq 1\}\cup \{e^n (fb)^n \mid n\geq 1\}
$$
is a FG language  but not a VPDA language.
 $L$ is generated by the FG grammar\\
$ S\to A\mid B \mid C \qquad A\to bAc\mid bc \qquad B\to fBd\mid fd \qquad C\to eCfb\mid efb $
\\
which has precedence relations $M$:
$$
b\doteq c, f\doteq d, e\doteq f, f\doteq b, b\lessdot b, f\lessdot f, e\lessdot e,  c\gtrdot c, d\gtrdot d, b\gtrdot f
$$
From $b^n c^n \subseteq L$ it follows $b$ must be a call and $c$ a return. For similar reasons, $f$ must be a call and $d$ a return. From $e^n (fb)^n \subseteq L$ it follows that at least one of $b$ and $f$ must be a return, a contradiction for a VP alphabet.
\end{proof}
\end{theorem}

\subsection*{FG with a partitioned precedence matrix}\label{SectRestrOPGareVPL}
We prove that the OPM structure obtained in the proof of Theor. \ref{TheorVPD2OPG} is a sufficient condition for an FG to generate a VPDA language thus obtaining a complete characterization of VPDA as a subclass of FG.
\\
For an alphabet $\Sigma$, let $M_T$ be an OPM such that there exists a partition of $\Sigma$ into three subsets $\Sigma_1, \Sigma_2$ and, $\Sigma_3$ satisfying the conditions:
\\
$ \forall a\in \Sigma_1,\forall  b\in \Sigma_1\cup\Sigma_3 : M_T[a,b]=\lessdot  $ and $ \forall a\in \Sigma_1,\forall b\in \Sigma_2 : M_T[a,b]=\dot=.$
\\
$ \forall a\in \Sigma_2,\forall  b\in \Sigma : M_T[a,b]=\gtrdot  $
\\
$ \forall a\in \Sigma_3,\forall  b\in \Sigma : M_T[a,b]=\gtrdot  $
\\
Then $M_T$ is termed a \emph{total VP-matrix}  representing the VP alphabet $\widehat{\Sigma} =\langle \Sigma_1, \Sigma_2,\Sigma_3\rangle= \langle \Sigma_c,\Sigma_r, \Sigma_i \rangle$, shown in Fig. \ref{figureTotalVPmatr}. Any OPM $M \subseteq M_T$ is termed a \emph{VP-matrix}.
\\
Observe that, for any  grammar $G$, such that $OPM(G)$ is a VP-matrix, any rule $A\to \alpha$ has $|\alpha|_{\Sigma}\leq 2$. The possible stencils (or skeletons) of the right parts of the rules are $NcN, NcNr, Nr, Ns$, and those obtained by erasing one or more $N$. Notice that the stencils $rN, crN$ are forbidden because $r$ does not yield precedence to any letter. It follows that, for any FG having a VP matrix, the length of any right part is $\leq 4$.
\begin{theorem}\label{TheorRestrOPGisVP}
Let $G$ be an FG such that $OPM(G)$ is a VP matrix. Then $L(G)$ is a VPDA language.
\begin{proof}
First we argue that the grammar generates any string in $L(G)$ with a syntax structure corresponding to the factorization presented in Lemma \ref{LemmaStruttVPL}. Then, in Lemma \ref{LemmaRestrOPG2VPA}, we construct a VPDA equivalent to $G$.
\\
Let $G$ satisfy the hypotheses of Theor. \ref{TheorRestrOPGisVP}. For every string $x\in L(G)$, the syntax tree induces the factorization\\
$ x = y c_0 z \qquad\text{  or  }\qquad x = y,  y=u_1 w_1 u_2 w_2 \ldots u_k w_k, z = v_1 c_1 v_2 c_2 \ldots c_{r-1} v_r $
\\where all terms are as
in Lemma \ref{LemmaStruttVPL}, and its syntax tree has the structure shown in Fig. \ref{figureTree}. It suffices to consider that the precedence relations of the VP matrix completely determine the skeleton of the syntax tree (see Fig. \ref{figureParsingRelations}).
\end{proof}
\end{theorem}

\begin{lemma}\label{LemmaRestrOPG2VPA}
Let $G=(\Sigma, V_N, P, S)$  satisfy the hypotheses of Theor. \ref{TheorRestrOPGisVP}. Then $L(G)$ is recognized by a VPDA automaton $\mathcal{A}=(\Sigma, Q, Q_0, \Gamma, \delta, Q_F)$, which can be effectively constructed.
 \begin{proof}
  We specify how to construct from the grammar rules a VPDA $\mathcal{A}$, that recognizes by final state and for convenience is nondeterministic. We recall the rule stencils are just the ones previously listed.
\\
We set $Q= V_N \cup \{q_0,p,q_F\}$, where  $q_0, p ,q_F \not\in V_N$. The pushdown vocabulary is
$$
\Gamma = \big( \left(V_N \cup \{-\} \right)  \times  \Sigma_c \times \left(V_N \cup \{-\} \right) \big) \cup \{\bot, Z_U\}
$$
Intuitively, $\mathcal{A}$ is built in such a way that it enters a state $B\in V_N$ after finishing the scanning of a substring syntactically rooted in $B$.
\\
In state $B$, reading a symbol  $c\in \Sigma_c$ (the only ones that yield precedence), $\mathcal{A}$ enters state $p$  and pushes  on the stack a symbol, for which two cases occur. The symbol is $Z_U$, if the $c$ is not to be matched by an $r$; it is $\langle B, c, C\rangle$, if the machine ``looks for'' a well-balanced string $w$ such that $C\stackrel{\ast}{\Rightarrow}w$. Simpler special cases also occur, such that $\mathcal{A}$ pushes on the stack a symbol $\langle B, c, -\rangle$ or $\langle -, c, -\rangle$, ``looking'' directly for $r$.
\\
In state $p$, reading a $c$,  $\mathcal{A}$ remains in the state and pushes on the stack either the symbol $Z_U$ if the $c$ is not to be matched, or a symbol $\langle -, c, C\rangle$ if it ``looks for'' a string $w$ such that $C\stackrel{\ast}{\Rightarrow}w$.
\\
Finally we describe the moves that read $r\in \Sigma_r$. If the stack is empty, the machine  enters a state $A$ associated to a nonterminal. If the top of stack is a symbol $\langle B, c, C\rangle$, the machine pops the stack and enters a state $A$. Here too some simpler special cases exist.
\\
The final states set is defined as $Q_F= \{A \mid S\stackrel{\ast}{\Rightarrow} \beta A, A\neq S \} \cup \{q_F\} \cup \{q_0 \text{ iff } S\to \varepsilon \in P\}$. Notice that a rule $A\to c B$ can be used only in a derivation  such as $S\stackrel{\ast}{\Rightarrow} \alpha A \Rightarrow \alpha c B \stackrel{\ast}{\Rightarrow} x$, otherwise $c$ would take precedence over some other letter. Thus, $A$ and $B$ are both in $Q_F$.

\begin{table}[h!]
\caption{\label{tableVPAMoves} Transition relation $\delta$ of $\mathcal{A}$.}
\renewcommand{\arraystretch}{1.3}
\begin{center}
\begin{tabular}{|l|p{4cm}|p{5cm}|}  \hline
  & rules  & $\delta$  \\
  \hline
 1&$A\to  s$ & $(q_0, s, A)$
 \\
  &$A\to  r$,\quad such that $S\stackrel{\ast}{\Rightarrow} A \alpha$  & $(q_0, r, \bot, A)$
  \\
  \hline
 2& $A\to s$ & $(p, s, A)$
 \\
 &  $A\to B s$ & $(B, s, A)$
 \\
 & $A\to  B r$ & $(B, r, \bot, A)$
  \\
  \hline
 3& $A\to c B$ & $(p, c, p, Z_U)$
  \\
 & $S\to B c C$ & $(B, c,p, Z_U)$
  \\
  \hline
 4& $S\to B c C r$ & $(B, c,p, \langle B,c,C \rangle )$
  \\
                 && $(C, r, \langle B,c,C \rangle, q_F)$
 \\
 & $S\to s$ & $(q_0, s,q_F)$
  \\
  & $S\to c$ & $(q_0, c,Z_U, q_F)$
  \\
 & $S\to r$ & $(q_0, r,\bot, q_F)$
  \\
  &$A\to B c C r$ & $(B, c, p, \langle B,c,C \rangle )$
  \\
                  && $(p, r, \langle B,c,C \rangle, A)$
  \\
  &$A\to B c  r$ & $(B, c, p, \langle B,c,- \rangle )$
  \\

                  && $(p, r,  \langle B,c,- \rangle, A)$
  \\
  \hline
  5& $A\to c B  r$ & $(p, c, p, \langle -,c,B \rangle )$
  \\
                  && $(B, r,  \langle -,c,B \rangle, A)$
  \\
  &  $A\to c r$ & $(p, c, p, \langle -,c,- \rangle )$
  \\
                  && $(B, r,  \langle -,c,- \rangle, A)$
  \\  \hline
\end{tabular}
\end{center}
\end{table}
The transition relation $\delta$ is then built from $P$ according to Table \ref{tableVPAMoves}. Notice that the derivations $S\stackrel{\ast}{\Rightarrow} A \alpha$ needed in section 1 of the table can be effectively computed.
\\
The proof of the equivalence $L(\mathcal{A})=L(G)$ somewhat mirrors the equivalence proof of Theor. \ref{TheorVPD2OPG}. For instance, from  section 2 of Table \ref{tableVPAMoves} the following lemma immediately descends:
$$
A\stackrel{\ast}{\Rightarrow}w, w \in \big( \Sigma_i \cup \Sigma_r\big)^\ast \iff \exists \sigma \in (\Gamma \setminus \{\bot\})^\ast , t \in Q \text{ such that } (t, \bot \sigma) \underset{w}{\stackrel*\mapsto} (A, \bot\sigma)
$$
Similarly, the lemma
$$
A\stackrel{\ast}{\Rightarrow}w, w \text{ well balanced } \iff \exists  \sigma \in (\Gamma \setminus \{\bot\})^\ast , t \in Q \text{ such that } (t, \bot \sigma) \underset{w}{\stackrel*\mapsto} (A, \bot\sigma)
$$
can be proved by a natural induction, taking as the basis the cases $A\to cr$ and $A\to s$, and then exploiting for the induction steps sections 2, 4, and 5  of Table \ref{tableVPAMoves}. Further details of the proof are omitted as fairly obvious.
\end{proof}
\end{lemma}
Second, we remark that various subclasses of VPDA languages recently considered correspond to restrictions on the VP-precedence matrix and/or on the stencils of the grammar rules. A nice illustration is the family BALAN \cite{Berstel:2001:BGT}. First, balanced grammars do not allow  any $c_i$ or $r_i$ to be unmatched. Thus an FG such that no rule has the stencils $Nc_iN, Nc_i, c_iN, c_i, Nr_i$ ensure the balancing property. Second, balanced grammars do not allow a $c_i$ to be matched by distinct returns $r_j,r_k$ (and similarly for $r_i$). An FG such that $|\Sigma_c|=|\Sigma_r|$ and the OPM submatrix identified by rows $\Sigma_{c_i}$ and columns $\Sigma_{r_i}$ contains $\dot=$ only on the diagonal, ensures the bijection of call and return letters.

\section{Closure properties}
All families considered here (except DPDA) share the property of being boolean algebras, for suitably defined subsets.  The core of the property dates back to the original approach by McNaughton and the "structure preserving" operations as in \cite{Crespi-ReghizziMM1978}. Other closure properties possessed by VPDA, though relevant and classical, have been less investigated. It appears that all the previous families more general than VPDA lack  some closure properties, as shown in the next table.
\begin{center}
\begin{tabular}{|c||p{4cm}|c|c|}\hline
   family  & boolean operations & concatenation, Kleene star & reversal  \\
  \hline\hline
  VPDA \cite{AluMad04}& yes for a fixed VP alphabet & yes for a fixed VP alphabet & yes \\ \hline
  FG & yes for compatible precedence matrices \cite{Crespi-ReghizziMM1978}& yes & yes (proved here) \\\hline
  HRDPDA & yes for H-synchronized languages \cite{conf/mfcs/NowotkaS07} & no \cite{caucal:DSP:2008:1743} & no (proved here)  \\
  \hline
\end{tabular}
\end{center}
The reversal of a FG language is generated by the specularly reversed rules; they are a FG grammar with a matrix obtained interchanging yield- and take-precedence relations.
\\
We observe that the boolean closure of  FG languages has been  proved in \cite{Crespi-ReghizziMM1978} by extending McNaughton's method for parenthesis languages. It states that the union of two FG having compatible precedence matrices is a FG language with compatible matrices, and similarly for the other operators. We notice that this is not implied by  the closure property \cite{conf/mfcs/NowotkaS07} of the equivalence class of H-synchronized HDPDA languages, although  two FG's having compatible matrices are necessarily H-synchronized\footnote{For brevity we omit the natural construction of the HDPDA equivalent to a FG grammar.}.
\\
On the other hand the closure of VPDA languages for a given VP alphabet, under the boolean operators and under reversal, are an immediate consequence of the same properties of the family of FG languages having compatible precedence relations,
\\
Since HRDPDA=RDPDA, the non-closure under reversal follows from a classical counterexample, used for proving the same for deterministic languages: the reversal of $\{1a^n b^n \mid n\geq 0\} \cup \{2a^n b^{2n} \mid n\geq 0\}$ is non-deterministic.
\\
The proof\footnote{A complete proof is under development.}  of concatenation and Kleene star closures for FG's is more intricate than with other traditional families of CF grammars due to the need to preserve the operator structure and the precedence relations. For kleene star the property requires the assumption that the $\dot=$ relation does not contain cycles.
\\
In conclusion, the FG family is currently the one, among the existing VPDA generalizations, that preserves the majority, and possibly the totality, of VPDA closure properties.

\section{Conclusions}\label{SectConclusion}
We mention some open questions raised by the present study.
\\
FG appears at present to be the family that preserves the majority, and possibly the totality, of VPDA closure properties, but we wonder whether more general families can be found with the same properties.
\par
In a different direction, it is possible to transfer to VPDA a rather surprising invariance property of FG. We recall the definition of \emph{Non-Counting context-free} grammar \cite{CreGuiMan78}, which extends the notion of NC regular language \cite{McNaughtPap71}. $L=L(G)$ is NC if for the parenthesized language $L(\widetilde{G})$, the following condition holds: $ \exists n>0: \forall x,v,w,\underline{v},y \in \Sigma^\ast $, where $w$ and $v w \underline{v}$ are well-parenthesized, and $\forall m\geq 0$, $x v^{n} w \underline{v}^{n} y \in \widetilde{L}$ if, and only if, $x v^{n+m} w \underline{v}^{n+m} y \in \widetilde{L}$. In general, two equivalent CF grammars may differ with respect to the NC property. However if an FG grammar is NC, then all equivalent FG grammars are NC \cite{CreGuiMan81}. Consider now, for a VPDA $L\subseteq \Sigma^{\ast}$, two equivalent VPDA recognizers. Notice the two VP alphabets may differ with respect to the 3-partition of the letters. The two corresponding FG's (Theor. \ref{TheorVPD2OPG}) may differ in precedence relations, but they are either both NC or both counting. We wonder whether such invariance property holds for other families of grammars generalizing VPDA.
\par
Last, it would be interesting to assess the suitability of Floyd languages for the applications that have motivated balanced grammars and VPDA. We observe that the greater generative capacity of FG's permits to define more realistic recursively nested structures. For instance, the VPDA approach uses single letters to represent a call $c$ and the corresponding return $r$, but this is just an abstraction. In real programming languages a call is a string typically containing the name of the invoked procedure and possibly a list of parameters. Also, at it is suggested by the example in the proof of Theor \ref{TheorStrictInclusion}, a return corresponding to a given call may use the same letters as some other call. This will cause conflicts in the partitioning of $\Sigma$, but can be dealt with by suitable precedence relations. Similar examples can be found in the area of mark-up languages.
\\
Finally, for application in model checking, the computational complexity of the decision problems for FG languages should be studied.

\bibliographystyle{plain}
\bibliography{VPDbib}

\begin{thebibliography}{10}

\bibitem{AluMad04}
R.~Alur and P.~Madhusudan.
\newblock Visibly pushdown languages.
\newblock In {\em STOC: ACM Symposium on Theory of Computing (STOC)}, 2004.

\bibitem{Berstel:2001:BGT}
J.~Berstel and L.~Boasson.
\newblock Balanced grammars and their languages.
\newblock In W.~Brauer et~al., editor, {\em Formal and Natural Computing},
  volume 2300 of {\em LNCS}, pages 3--25. Springer, 2002.

\bibitem{conf/dlt/Caucal06}
D.~Caucal.
\newblock Synchronization of pushdown automata.
\newblock In O.~H. Ibarra and Z.~Dang, editors, {\em Developments in Language
  Theory}, volume 4036 of {\em LNCS}, pages 120--132. Springer, 2006.

\bibitem{caucal:DSP:2008:1743}
D.~Caucal.
\newblock Boolean algebras of unambiguous context-free languages.
\newblock In R.~Hariharan, M.~Mukund, and V.~Vinay, editors, {\em IARCS Annual
  Conference on Foundations of Software Technology and Theoretical Computer
  Science (FSTTCS 2008)}, Dagstuhl, Germany, 2008. Schloss Dagstuhl -
  Leibniz-Zentrum fuer Informatik, Germany.

\bibitem{conf/csr/CaucalH08}
D.~Caucal and S.~Hassen.
\newblock Synchronization of grammars.
\newblock In Edward~A. Hirsch, Alexander~A. Razborov, Alexei~L. Semenov, and
  Anatol Slissenko, editors, {\em CSR}, volume 5010 of {\em LNCS}, pages
  110--121. Springer, 2008.

\bibitem{crespi-72}
S.~Crespi-Reghizzi.
\newblock An effective model for grammar inference.
\newblock In {\em Information Processing 71}, pages 524--529. North-Holland
  Publishing Co., 1972.

\bibitem{CreGuiMan78}
S.~Crespi-Reghizzi, G.~Guida, and D.~Mandrioli.
\newblock Noncounting context-free languages.
\newblock {\em JACM: Journ. of the ACM}, 25:571--580, 1978.

\bibitem{CreGuiMan81}
S.~Crespi-Reghizzi, G.~Guida, and D.~Mandrioli.
\newblock Operator precedence grammars and the noncounting property.
\newblock {\em SICOMP: SIAM Journ. on Computing}, 10:174---191, 1981.

\bibitem{Crespi-ReghizziMM1978}
S.~Crespi-Reghizzi, D.~Mandrioli, and D.~F. Martin.
\newblock Algebraic properties of operator precedence languages.
\newblock {\em Information and Control}, 37(2):115--133, May 1978.

\bibitem{Fischer69}
M.~J. Fischer.
\newblock Some properties of precedence languages.
\newblock In {\em STOC '69: Proc. first annual ACM symp.m on Theory of
  computing}, pages 181--190, New York, NY, USA, 1969. ACM.

\bibitem{FisPnu01}
D.~Fisman and A.~Pnueli.
\newblock Beyond regular model checking.
\newblock {\em FSTTCS: Foundations of Software Technology and Theoretical
  Computer Science}, 21, 2001.

\bibitem{Floyd1963}
R.~W. Floyd.
\newblock Syntactic analysis and operator precedence.
\newblock {\em J. ACM}, 10(3):316--333, 1963.

\bibitem{GruneJacobs:08}
D.~Grune and C.~J. Jacobs.
\newblock {\em Parsing techniques: a practical guide}.
\newblock Springer, New York, 2008.

\bibitem{Harrison78}
M.~A. Harrison.
\newblock {\em Introduction to Formal Language Theory}.
\newblock Addison Wesley, 1978.

\bibitem{hopullman:automata}
J.~E. Hopcroft and J.~D. Ullman.
\newblock {\em Introduction to Automata and Formal Languages}.
\newblock Addison-Wesley, Reading, MA, 1979.

\bibitem{conf/csr/LimayeMM08}
N.~Limaye, M.~Mahajan, and A.~Meyer.
\newblock On the complexity of membership and counting in height-deterministic
  pushdown automata.
\newblock In E.A.~Hirsch et~al., editor, {\em CSR}, volume 5010 of {\em LNCS},
  pages 240--251. Springer, 2008.

\bibitem{McNaughton67}
R.~McNaughton.
\newblock Parenthesis grammars.
\newblock {\em J. ACM}, 14(3):490--500, 1967.

\bibitem{conf/mfcs/NowotkaS07}
D.~Nowotka and J.~Srba.
\newblock Height-deterministic pushdown automata.
\newblock In Ludek Kucera and Anton{\'i}n Kucera, editors, {\em MFCS}, volume
  4708 of {\em LNCS}, pages 125--134. Springer, 2007.

\bibitem{McNaughtPap71}
{R. McNaughton} and {S. Papert}.
\newblock {\em Counter-free Automata}.
\newblock {MIT} {P}ress, {C}ambridge, {USA}, 1971.

\bibitem{Tha67}
J.~W. Thatcher.
\newblock Characterizing derivation trees of context-free grammars through a
  generalization of finite automata theory.
\newblock {\em Journ. of Comp. and Syst.Sc.}, 1:317--322, 1967.

\bibitem{LaTorreNP06}
S.~La Torre, M.~Napoli, and M.~Parente.
\newblock On the membership problem for visibly pushdown languages.
\newblock In S.~Graf and W.~Zhang, editors, {\em ATVA {Automated Technology for
  Verification and Analysis}}, volume 4218 of {\em LNCS}, pages 96--109.
  Springer, 2006.

\end{thebibliography}
\end{document}